\documentclass[aps,prl,twocolumn,groupedaddress]{revtex4-2}

\usepackage{graphicx}
\usepackage{natbib,hyperref}
\usepackage[english]{babel}

\begin{document}

\title{Towards skyrmion--superconductor hybrid systems}

\author{Andr\'e~Kubetzka}
\author{Jan~M.~B\"urger}
\author{Roland~Wiesendanger}
\author{Kirsten~von~Bergmann}
\email[]{kbergman@physnet.uni-hamburg.de}
\affiliation{Department of Physics, University of Hamburg, Jungiusstr.~11, 20355 Hamburg, Germany}

\date{\today}

\begin{abstract}
Spin-polarized scanning tunneling microscopy is used to identify the magnetic state of different thin films on a Re(0001) substrate, which becomes superconducting below 1.7\,K. All magnetic films contain an Fe/Ir interface, which is known to facilitate the emergence of non-collinear magnetic order. For Fe monolayers on ultra-thin Ir films of different thickness we find several different atomic-scale magnetic states. For Pd/Fe bilayers we find nano-scale spin spirals as magnetic ground states for Ir thicknesses of three and four atomic layers. In applied magnetic fields skyrmions emerge, and in remanence non-trivial magnetic textures survive. This demonstrates the possibility to prepare skryrmion-hosting magnetic films on superconducting substrates.
\end{abstract}

\maketitle

Recently, several theoretical studies highlighted interesting properties that are predicted to arise when magnetic skyrmions are in direct contact to a superconductor~\cite{nakosaiPRB2013,chenPRB2015,yangPRB2016,pershogubaPRB2016,gungorduPRB2018,rexPRB2019,garnierCP2019}. From a practical point of view the realization of hybrid skyrmion--superconductor systems is experimentally not trivial: on one hand the formation of magnetic skyrmions occurs usually only in a limited phase space, defined by a subtle balance of several different material-specific magnetic interactions and energies, in particular the isotropic exchange interaction, the Dzyaloshinskii-Moriya interaction, the magnetocrystalline anisotropy energy, and the Zeeman energy~\cite{nagaosaNN2013}. On the other hand, the number of superconducting materials which can be interfaced with a skyrmion-hosting magnetic film is limited. To make things worse, magnetic skyrmions typically arise upon application of an external magnetic field, which is detrimental to superconductivity. 

The Fe/Ir interface has proven to be beneficial for magnetic skyrmion formation~\cite{heinzeNP2011,rommingS2013,hsuNN2017,soumyanarayananNM2017,hsuNC2018}. In the fcc-stacked Fe monolayer on Ir(111) a square nanoskyrmion lattice emerges with about 14 atoms in the magnetic unit cell~\cite{heinzeNP2011}. It is driven by the competition of exchange interactions, the Dzyaloshinskii-Moriya interaction, and higher-order magnetic interactions, and exists also in the absence of an external magnetic field. Indeed, it does not react to applied magnetic fields of up to $B = 9$\,T~\cite{vonbergmannJPCM2014}. The hcp-stacked Fe monolayer on Ir(111) exhibits a hexagonal nanoskyrmion lattice~\cite{vonbergmannNL2015}. Other magnetic films with Fe/Ir interfaces are Pd/Fe atomic bilayers~\cite{rommingS2013,kubetzkaPRB2017}, the Fe triple layer~\cite{hsuNN2017}, and the hydrogenated Fe double layer on Ir(111)~\cite{hsuNC2018}, all of which exhibit a nano-scale spin spiral ground state with skyrmions induced by external magnetic fields. In addition, {Ir/Fe/Co/Pt} multilayers~\cite{soumyanarayananNM2017} have shown skyrmions. 

At first glance it appears promising to interface Fe with a different 5$d$ single crystal which is superconducting, such as Re(0001) with a critical temperature of $T = 1.7$\,K. The Re(0001) surface can be prepared to be very clean as demonstrated by scanning tunneling microscopy (STM) measurements~\cite{ouaziSS2014}, and the superconducting properties can be investigated with tunnel spectroscopy~\cite{kimSA2018,schneidernQM2019}. However, a pseudomorphic monolayer of Fe on Re(0001) exhibits strong antiferromagnetic coupling; on a hexagonal lattice this gives rise to the Néel state with $120^\circ$ between all nearest neighbor magnetic moment pairs~\cite{hardratPRB2009,ouaziPRL2014}. In contrast to Fe/Ir, no magnetic skyrmions were observed in Fe-based ultrathin films directly on the Re(0001) substrate. One experimental example of a magnetic film with magnetic skyrmions in remanence, epitaxially grown on a superconductor, is a Co monolayer on Ru(0001)~\cite{herveNC2018}; the skyrmions are comparably large with diameters of some ten nanometers and the critical temperature for superconductivity is $0.5$\,K, even lower than that of Re(0001). 
 
Here, we demonstrate a route towards a hybrid skyrmion-superconductor model-type system, by preparing the well-known skyrmion-friendly interface Fe/Ir on a superconducting Re(0001) single crystal substrate. These materials are a promising match because the lateral lattice mismatch between the hexagonal surfaces of Ir(111) and Re(0001) is only $2\%$ and we anticipate pseudomorphic growth of ultrathin Ir films on a Re(0001) substrate. We employ epitaxial growth to first cover a superconducting Re(0001) crystal with atomically thin Ir layers, thereby mimicing the substrate for skyrmion hosting magnetic films. Then we create an Fe/Ir interface by adding submonolayer amounts of Fe; subsequently we also deposit Pd to fine tune the magnetic properties. We show that both the magnetic states of the Fe monolayers as well as Pd/Fe atomic bilayers depend critically on the thickness of the Ir film and the stacking of the adlayers. 

Using (spin-polarized) STM~\cite{SOM} we study the growth and the magnetic states of various samples with different amounts of Pd,Fe,Ir on a Re(0001) substrate. We find that post-annealing of Ir/Re samples is beneficial for the formation of extended areas of Ir (see Supplementary Information~\cite{SOM}). Ir grows pseudomorphic up to a thickness of at least 5 atomic layers (AL); while we cannot derive information about the relative stacking of each layer, we see no sign of a coexistence of both possible stackings within any of the layers. We observe the formation of islands, and typically several different layer thicknesses of Ir coexist on one sample~\cite{SOM}. 

\begin{figure*}[b]
\includegraphics{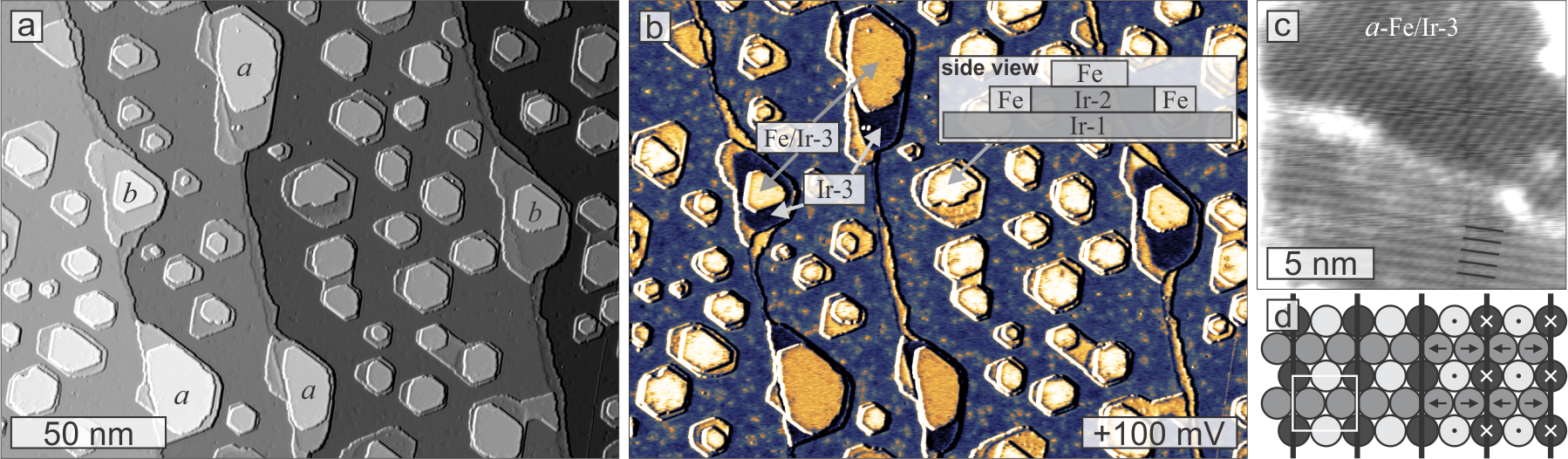}
\caption{Growth and magnetism of Fe on Ir on Re(0001). \textbf{(a)}~STM constant-current image (partially differentiated) of a sample with about $0.3$\,AL of Fe deposited onto a sample with about $1.3$\,AL of Ir on Re(0001). The letters $a$ and $b$ indicate different types of Fe/Ir-3 islands. \textbf{(b)}~Simultaneously acquired $dI/dU$~map, in which the different layer compositions can be discriminated. \textbf{(c)}~STM constant-current of an $a$-Fe/Ir-3 area; the stripes are of magnetic origin, gray-scale is $20$\,pm from black to white. \textbf{(d)}~Sketch of the magnetic unit cell (white rectangle) together with a tentative magnetic structure. Measurement parameters: (a),(b) $U = +100$\,mV, $I = 1$\,nA; (c) $U = +5$\,mV, $I = 2$\,nA; all: $T = 4.2$\,K, Cr tip.}
\label{fig1}
\end{figure*}

When about $1.3$\,AL of Ir are deposited onto Re(0001) and subsequently annealed, we find a fully closed Ir monolayer \mbox{(Ir-1)} with many single atomic layer Ir islands on top, which then have the local coverage of two atomic Ir layers \mbox{(Ir-2)}; some Ir-3 layers are also observed. After subsequent deposition of about $0.3$\,~AL of Fe such a sample has the morphology as shown in the STM constant-current image of Fig.\,\ref{fig1}(a). Together with the simultaneously measured $dI/dU$~map in Fig.\,\ref{fig1}(b), we find that Fe dominantly forms monolayer high islands on Ir-2, and also attaches to the rim of Ir-2 islands, resulting in Fe/Ir-2 and \mbox{Fe/Ir-1} patches, respectively (see side view sketch in (b)). For the Fe monolayers both on Fe-1 and on Fe-2 we find a $(\sqrt{3}\times\sqrt{3})\rm{R}30^\circ$ magnetic superstructure~\cite{SOM}, indicating that the magnetic ground state is the Néel state, which is also the ground state of the Ir-free Fe/Re(0001)~\cite{ouaziPRL2014}.

On the Ir-3 islands, two different types of pseudomorphic Fe monolayers can be found, as indicated by $a$ and $b$ in the STM constant-current image shown in Fig.\,\ref{fig1}(a); they can clearly be distinguished by their $dI/dU$~signals, see Fig.\,\ref{fig1}(b). We conclude that the difference in the two Fe/Ir-3 layers is the stacking of the Fe, i.e.\ one of them grows in fcc and the other one in hcp configuration with respect to the underlying Ir layers. A closer view of an \mbox{$a$-Fe/Ir-3} island is shown in the STM constant-current image of Fig.\,\ref{fig1}(c); the stripes are of magnetic origin and two rotational domains are present. The stripes run perpendicular to close-packed atomic rows and the spacing between the stripes is about $0.58$\,nm, which is roughly twice the nearest neighbor Re distance of $0.2761$\,nm. The derived magnetic unit cell is sketched in Fig.\,\ref{fig1}(d) together with a suggestion for the magnetic state.

\begin{figure}
\centering
\includegraphics{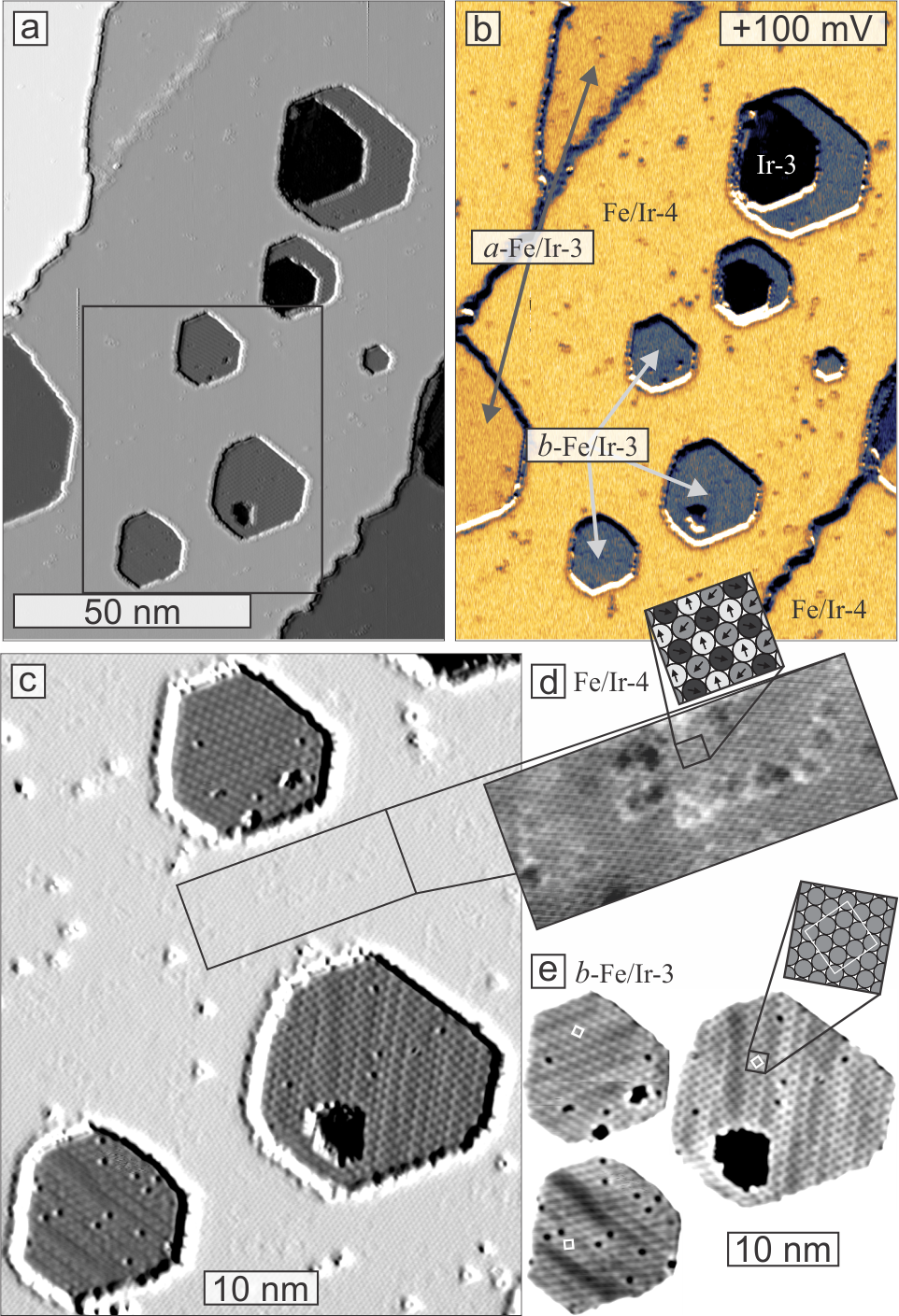}
\caption{Growth and magnetism of $b$-Fe/Ir-3 and Fe/Ir-4 on Re(0001). \textbf{(a)}~STM constant-current image (partially differentiated) of a sample with about $0.9$\,AL of Fe deposited onto a sample with about $3.6$\,AL of Ir on Re(0001); the Ir/Re sample before deposition of Fe is shown in Suppl.\,Fig.\,S2(c),(d)~\cite{SOM}. \textbf{(b)}~Simultaneously acquired $dI/dU$~map, in which the different layer compositions can be discriminated. \textbf{(c)}~STM constant-current image (partially differentiated) of the area indicated by the box in (a). \textbf{(d)}~STM constant-current image of the indicated Fe/Ir-4 area exhibiting the N\'eel state; gray-scale is $10$\,pm from black to white. \textbf{(e)}~STM constant-current images of the three different $b$-Fe/Ir-3 areas of (c), which exhibit all possible rotational domains of an atomic-scale roughly square magnetic state; gray-scale is $30$\,pm from black to white. Measurement parameters: (a),(b) $U = +100$\,mV, $I = 1$\,nA; (c)-(e) $U = +5$\,mV, $I = 4$\,nA; all: $T = 4.2$\,K, Cr-tip.}
\label{fig2}
\end{figure}

Figure\,\ref{fig2} shows a different sample with a nearly complete AL of Fe on Ir-3 and Ir-4 layers. We find that $b$-Fe/Ir-3 is stabilized in vacancy islands within the Ir-4 layer, whereas the $a$-Fe is typically found near buried step edges of the Re substrate. Again, the different stackings can be distinguished by their $dI/dU$~signal, see Fig.\,\ref{fig2}(b). The area indicated by the box in (a) is shown in (c); to enhance the magnetic superstructures this STM constant-current image has been partially differentiated. Figure\,\ref{fig2}(d) shows an enlarged view of the indicated Fe/Ir-4 area. The observed $(\sqrt{3}\times\sqrt{3})\rm{R}30^\circ$ magnetic superstructure is characteristic of the N\'eel state, see inset; it occurs in two inverted domains, as seen on the right and left side of this area. Figure\,\ref{fig2}(e) shows the three $b$-Fe/Ir-3 areas, in which a different magnetic superstructure occurs. The magnetic state is roughly square with a lattice constant of about $0.77$\,nm, see inset; the three areas show the three possible rotational domains. The uni-directional beating pattern possibly originates from confinement effects of the magnetic state due to the boundary, or from the incommensurability of the magnetic state with respect to the atomic lattice. This magnetic superstructure is reminiscent of the nanoskyrmion lattice found in fcc-stacked Fe on Ir(111)~\cite{heinzeNP2011}, with a slightly reduced ($10\%$) lattice constant. This suggests that the $b$-Fe/Ir-3/Re(0001) is another example of an atomic-scale skyrmion lattice, in this case in direct vicinity to a superconducting material. 

\begin{figure}
\includegraphics{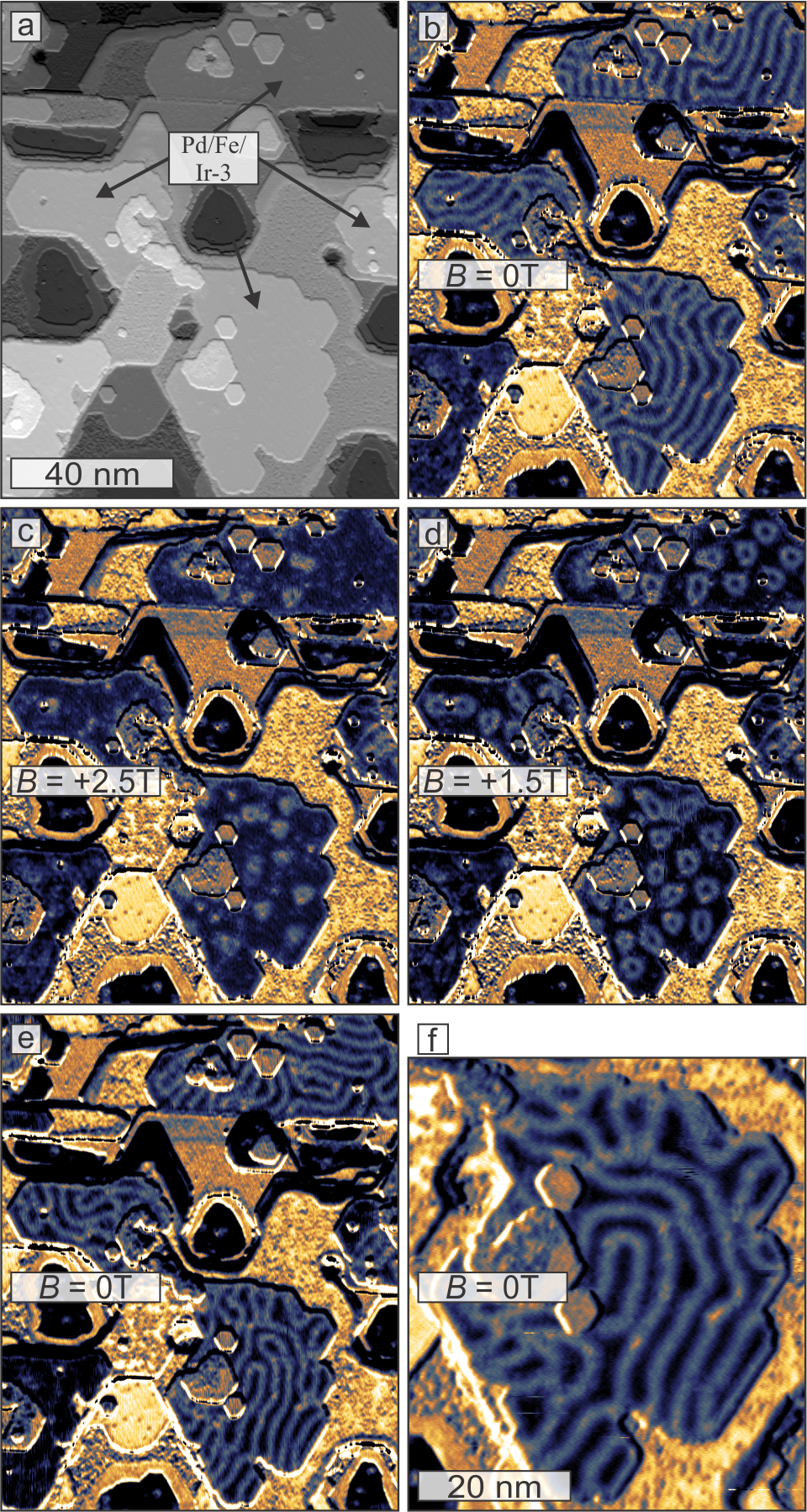}
\caption{Magnetic field dependence of Pd/Fe/Ir-3 on Re(0001). \textbf{(a)}~STM constant-current image (partially differentiated) of a sample with about $0.5$\,AL of Pd on $0.9$\,AL of Fe on $2.6$\,AL of Ir on Re(0001). \textbf{(b)}~$dI/dU$ map of the same area in the magnetic virgin state. \textbf{(c)},\textbf{(d)},\textbf{(e)}~$dI/dU$~maps at the indicated magnetic fields, in chronological order. \textbf{(f)}~$dI/dU$~map of the central bottom Pd/Fe/Ir-3 island in a different magnetic state in remanence. Measurement parameters: $U = -40$\,nA, $I = 1$\,nA (except for (f) with $I = 2$\,nA), $T = 8.6$\,K, Fe/W-tip without a sign of spin-polarization in this measurement.}
\label{fig3}
\end{figure}

In order to utilize possibly emerging exotic phenomena like Majorana bound states for, e.g., quantum computation, it is desirable to be able to manipulate and control the magnetic state. This usually proves to be difficult for atomic-scale periodic magnetic structures: whereas it has been shown that the square nanoskyrmion lattice can switch between its different rotational domains~\cite{vonbergmannPRB2012}, the isolation and manipulation of a single skyrmion has not been possible. However, previous investigations have shown that the atomic-scale magnetic nanoskyrmion states of the Fe monolayers on Ir(111) can be tuned into magnetic-field induced skyrmion systems by Pd overlayers~\cite{rommingS2013}. Such Pd/Fe atomic bilayers on Ir(111) can stabilize not only skyrmion lattices, but also metastable single magnetic skyrmions that can be created and annihilated with the tip of an STM~\cite{rommingS2013,kubetzkaPRB2017}. In addition, it has been shown that these individual magnetic skyrmions can be moved laterally across the film by coupling them to adsorbed clusters, which can be moved across the surface by manipulation with the STM tip~\cite{hannekenNJP2016}. In this manner a braiding of skyrmion-induced Majorana bound states might be possible.

To find a system in which isolated magnetic skyrmions can occur, we have prepared Pd/Fe atomic bilayers on Ir films on Re(0001). Figure\,\ref{fig3} shows a preparation with about half an atomic layer of Pd on a sample with nearly a full atomic layer of Fe on about $2.6$\,AL of Ir on Re(0001). Comparison of the STM constant-current image (a) and the simultaneously aquired $dI/dU$~map (b) allows an assignment of the layer sequences of the exposed layers, see also~\cite{SOM}. We find that the Pd/Fe atomic bilayer areas on Ir-3 exhibit a stripe pattern. When an external magnetic field of $B=+2.5$\,T is applied, see Fig.\,\ref{fig3}(c), the stripes vanish and instead round objects emerge. At $B=+1.5$\,T bright rings are observed, Fig.\,\ref{fig3}(d). This feature is characteristic of magnetic skyrmions imaged by the tunnel non-collinear magnetoresistance effect (NCMR)~\cite{hannekenNN2015,crumNC2015,kubetzkaPRB2017}, which is based on the fact that the electronic states of locally non-collinear spin textures are different from those of magnetically collinear configurations. Due to this difference also the vacuum density of states, which is the parameter determining the signal in STM, changes locally across a skyrmion. The NCMR has been identified in Pd/Fe/Ir(111)~\cite{hannekenNN2015,kubetzkaPRB2017} and we find similar characteristics in the energy-resolved differential conductance of Pd/Fe/Ir-3, as measured by tunnel spectrocopy~\cite{SOM}. In analogy to Pd/Fe/Ir(111), we conclude that the magnetic ground state of Pd/Fe/Ir-3 is a spin spiral with a period of roughly $7.5$\,nm, which in NCMR images appears as stripes with distances of half the magnetic period~\cite{SOM}. The magnetic field induces magnetic skyrmions that are imaged as rings at intermediate magnetic fields and transform into dots at higher magnetic fields~\cite{hannekenNN2015,kubetzkaPRB2017}. 

The magnetic skyrmions in the Pd/Fe film, Fig.\,\ref{fig3}(c,d), are induced by applied magnetic fields which are significantly larger than the critical magnetic field for the superconductivity of Re(0001). However, it has been shown that zero-field skyrmions can exist in thin-film systems, either due to confinement~\cite{boulleNN2016,hoPRA2019}, in remanence~\cite{pollardNC2017,carettaNN2018}, or due to a favorable interplay of magnetic interactions~\cite{meyerNC2019}. Indeed we find that both in the virgin state (b) and in the remanent state (e) the magnetic state of Pd/Fe/Ir-3 has several areas, which are expected to give rise to local topological charges~\cite{cortes-ortunoPRB2019}, such as branches, loose ends, or expanded skyrmions. The details of the local spin texture can be changed, as evident by imaging the same area multiple times, compare Fig.\,\ref{fig3}(e) and (f) (see also~\cite{SOM}). In this way different magnetic states can be prepared within the identical sample area, which allows a disentanglement of the impact of the magnetic texture on exotic properties from the influence of structural or electronic effects. 

In conclusion, we have demonstrated that the Fe/Ir interface can be prepared on a Re substrate, where it still facilitates the formation of magnetic skyrmions. The Fe monolayer on Ir(111) is characterized by a competition of different magnetic interactions and energies, and this is reflected by the sensitivity of its magnetic state to the stacking and number of supporting Ir layers on Re(0001). Whereas Fe on Ir-1, Ir-2 and Ir-4 exhibit the N\'eel state, on \mbox{Ir-3} the two stackings of Fe show two different magnetic structures: a 0.58\,nm spin spiral and a 0.77\,nm skyrmion lattice. The addition of a Pd overlayer onto Fe/Ir-3 and Fe/Ir-4 leads to spin spiral ground states with periods on the order of 7\,nm. Application of external magnetic fields results in the formation of magnetic skyrmions. Also in zero field magnetic objects are observed, which are expected to have non-vanishing topological charge. Both the atomic-scale skyrmion lattice and the zero-field magnetic objects are in direct vicinity to the Re substrate, which could serve as a platform to search for exotic phenomena due to the interplay of magnetic skyrmions and superconductors below the critical temperature.

\begin{acknowledgments}
K.v.B. and A.K. acknowledge funding by the Deutsche Forschungsgemeinschaft (DFG, German Reseach Foundation) - 402843438; 408119516, 418425860. R.W. acknowledges financial support by the European Union via the ERC Advanced Grant ADMIRE.
\end{acknowledgments}

\end{document}